# Clickbait in Education – Positive or Negative? Machine Learning Answers

Adil Rajput[1]

[1] College of Engineering Effat University, Jeddah Kingdom of Saudi Arabia

**Abstract.** The topic of clickbait has garnered lot of attention since the advent of social media. Meriam-Webster defines Clickbait as something designed to make readers want to click on a hyperlink especially when the link leads to content of dubious value or interest. Clickbait is used synonymously with terms with negative connotations such as yellow journalism, tabloid news etc. Majority of the work in this area has focused on detecting clickbait to stop being presented to the reader. In this work, we look at clickbait in the field of education with emphasis on educational videos that are authored by individual authors without any institutional backing. Such videos can become quite popular with different audiences and are not verified by any expert. We present findings that despite the negative connotation associated with clickbait, audiences value content regardless of the clickbait techniques and have an overall favorable impression. We also establish initial metrics that can be used to gauge the likeness factor for such educational videos/MOOCs.

## 1. Introduction

Clickbait has gained attention of researchers for a long time and advent of social media has intensified the debate further. In simple terms, the goal of click bait is to attract attention of the users by attractive text or picture or both. Many researchers consider the term clickbait synonymous to yellow journalism where the content is sensationalized without providing any well-researched facts. Yellow Journalism has had a negative connotation as the mainstream population welcomes well-researched news in a society. The attitude towards clickbait is akin to that of yellow journalism as many authors have proposed ways to avoid clickbait on the net [3].

While traditional media such as print media, television etc. would attract mainly corporations who could be able to afford such advertising media, the Internet in general and search engines allowed regular users to advertise for their services as well. The social media experience amplified the model even further resulting in a network that would allow a wave of global advertisements. Machine Learning, Big Data and Natural Language Processing took the model a step further where users would be targeted per their choices on the Internet (both from a commercial and entertainment perspective).

Authors in [2] discuss the impact of clickbait in education and mention the case of an academic who garnered massive citations due to the controversial nature of colonialism and why the author in [4] believed that colonialism got a bad reputation. Critics claimed that the author used a controversial topic to gain extra attention in a quick fashion and was akin to the techniques used in clickbait/yellow journalism. This begs the following question: Does clickbait exist in education? Academic institutions globally have always made claims to attract students such as "Quality Education", "Better Job prospects" etc. However, we could not come across any study that would denounce such behavior as is the case for other industries.

The open courseware project announced by MIT about 15 years ago proved to massively popular with both active and lifelong learning students. However, in this case the reputation of the academic institution backed up the material placed online and gained instant credibility. This ushered an era of Massive Online Open Courses (MOOCs) and educational videos made popular through various social media channels especially YouTube. Various training classes that were traditionally expensive and required physical presence and at times required expensive setup were posted on YouTube – mostly in the hopes of monetizing the knowledge that various people possessed. Given the study in [26], authors of such videos would divide the training material into shorter segments to ensure that users maximize the benefit from such videos. The authors advertise for their educational content by 1) either embedding the link to other videos or their channel in the Description field that can serve content on various topics or 2) at times embedding it in the comments section of their own videos. In both the cases, the author writes text (and images in some cases) that would be enticing to end users. Examples include "Learn to hack into a network", "Learn Python in an hour" etc. The techniques used fulfill all the criteria of clickbait as the majority of the authors make claims that were neither well researched nor verifiable in an objective fashion (at least not easily).

In this paper, we aim to evaluate whether the clickbait model in social media is necessarily perceived as negative. Specifically, we present the following:

1. Briefly establish the case that clickbait has a negative connotation (Section 2 – Literature Review)
2. Establish the metrics that we use to gauge the interest of viewership (Section 3 – Experimental Setup)
3. Results of our evaluation (Section 4 – Results and Discussion)

## 2.   Literature Review

Lowenstein in [6] argues about the psychological factor that underlies the theory of clickbait. Specifically, the author argues about the information-gap of the reader/viewer that would propel the person to click on the underlying link as the title would prove to be enticing to the reader/viewer and they would click on it to satisfy the curiosity. Authors in [7] study listicles from Buzzfeed and notice that 85% use a cardinal number and employ the use of strong nouns. Authors in [8] establish that the use of discourse deixis and cataphora is widely common in ad-centric clickbait links. Specifically, cataphora in linguistic refers to pointing to something later in the text. As an example, "When he arrived, Smith smiled". In this sentence the pronoun he defers the mentioning of Smith to later part of the sentence engaging the reader. Discourse Deixis means using part of a text/speech to refer to part of the utterance that came or has yet to come. As an example, consider the statement "This story will leave you mesmerized". In this sentence, the speaker/writer is referring to a part of speech/text that has not come yet. Authors in [9] develop a corpus by gathering about 3000 popular tweets and present a model on detecting clickbait links. Specifically, the study gathered data on 215 features and divide the data into three categories namely 1) the teaser message 2) the linked webpage and 3) meta information. The authors use machine learning algorithms and present a novel technique to detect click baits. Authors in [10] refer to clickbait as tabloid journalism and deem it as a form of deception. This reiterates the work of this paper where majority of the literature (if not all) consider clickbait as a negative tactic. The authors review both textual and non-textual techniques to detect clickbait and establish that a hybrid approach is the best way to detect such links. In [11], Silverman contends that tabloid journalism is a detriment to professional journalism. Authors in [12] tackle the topic of verification of news item and discuss the methodological limitations of such techniques. [15] describes an initiative launched in 2017 where a challenge was launched to see how clickbait can be detected in social media. Based on this initiative, [13] take the efforts against clickbait to another level and establish the need of a corpus that can be utilized to detect clickbait. The topic of corpus has garnered lot of attention in various field as authors in [20] also use twitter to build a corpus to detect mental health symptoms. In [14], the authors build a browser extension that can help detect and block clickbait links. [16] describes the solution of Zingel Clickbait detector which evaluates each tweet for the presence of clickbait. Please note that majority of the techniques rely upon Natural Language Processing techniques (NLP) and such techniques' application goes beyond into various areas such as sentiment analysis etc. [21]. As opposed to the stance by the aforementioned work, authors in [17] do talk about the positive aspect of clickbait in academic articles. The authors limit the work in Frontiers in Psychology and the authors own articles and contend that positive framing of an academic article and avoiding playing with words result in a positive dissemination of the work. The authors confirmed the findings with their own articles. However, the result of authors is specific to text articles and does not take academic/educational videos into account. Furthermore, the articles used by the authors are mostly refereed and hence the reader has some level of confidence when reading such articles. Our work looks into the topic of education videos specifically and none of the videos we looked at have been evaluated by experts in the subject field.

Learning Analytics as a field is not new as researchers have always been concerned about gathering data about learners to help optimize the delivery of course material. Authors in [22] discuss the importance of establishing the teacher as a mentor and reevaluate the passive mode in which classes are being taught. They look at ways where technology can help in this vein. The work done in [23] proposes a conceptual framework akin to a maturity model focused on integrating social network research with importance of technology and explores how the concept of smart education has evolved. Furthermore, the work done in [27-28] takes these concepts and focuses on establishing metrics and key performance indicators that would provide feedback to gauge the effectiveness of smart education.

## 3.     Experimental Setup and Metrics

We conducted our experiments by writing crawlers that would scavenge YouTube for a particular topic. Specifically, given the attention cybersecurity has garnered in the lately, we chose the following two terms "Replay Attacks" and "MITM attacks" (Man in the Middle Attacks)

### 3.1.    Metrics Used

The problem faced during the experiments was to establish metrics that can serve a proxy to the feedback provided by the user. Many studies have used the "Like" and "Dislike" feature available on YouTube to reflect the sentiments of the user. However, not all the users who view a video would necessarily take the initiative to click on the like/dislike button. Another factor that we considered was the comment section where viewers can describe their opinion about the content being provided. Based on this, we came up with the following metrics to serve a proxy to user feedback.

- Number of Views
- Number of Likes
- Number of Dislikes
- Likes to Dislikes ratio (Higher ratio reflects a well-liked content)
- Number of comments (Higher number indicates a well-liked or not-liked all video at all as user will only go out of their way generally to write a comment in such cases)

### 3.2.    Experimental Setup

For this paper, we used the Python language along with the API provided by YouTube to collect the requisite data and perform a depth-first search approach to collect the metrics. Specifically, we do the following:

1. Search for videos under the titles "Replay Attacks" and "MITM attacks". For each of the video, we collect the aforementioned metrics
2. We also check the description field and see if the author has embedded a link to another YouTube video
3. If the video is also authored by the same author, we repeat steps 1 and 2

Once we collected the data, we wanted to see if the clickbait technique (embedding links in the description field) turns away the users as is the perception of clickbait or the viewers become more interested in the topic at hand.

## 4. Results and Discussion

For this test, searched for ten videos initially and performed a depth-first search algorithm as described above. One of the underlying assumptions of the search is that the popular videos will come up first as YouTube rewards viewership. However, what is not obvious right away from such a search is whether the author of a particular video would continue to provide high quality videos or otherwise. The following tables summarizes the results for two videos along with their embedded links during our search. The first row of the table indicates the initial video that came up in our search. Subsequent rows show the number of embedded videos discovered by our crawler under the clickbait model described above.

| Views | Likes | Dislikes | Comments | Likes/Dislikes | Comments/Views |
|---|---|---|---|---|---|
| 34343 | 97 | 5 | 5 | 24.25 | 6868.6 |
| 12432 | 441 | 19 | 35 | 23.21 | 355.2 |
| 18506 | 76 | 12 | 4 | 6.33 | 19 |
| 3607 | 11 | 1 | 1 | 11 | 11 |
| 573 | 17 | 0 | 8 | 10 | 2.125 |

Table 1: Video 1 Statistics

| Views | Likes | Dislikes | Comments | Likes/Dislikes | Comments/Views |
|---|---|---|---|---|---|
| 30164 | 98 | 1 | 1 | 98 | 98 |
| 987 | 85 | 0 | 3 | 10 | 329 |
| 27684 | 138 | 2 | 0 | 69 | 0 |
| 18490 | 266 | 6 | 25 | 44.33 | 10.64 |
| 4403 | 4403 | 5 | 61 | 880.6 | 72.18 |

Table 2: Video 2 Statistics

Based on the data from performing a depth first search on 10 videos for two topics, we obtained the following average .

| Views | Likes | Dislikes | Comments | Likes/Dislikes | Comments/Views |
|---|---|---|---|---|---|
| 25034 | 325 | 18 | 53 | 18.05 | 472.4 |
| 17345 | 165 | 21 | 33 | 23.21 | 525.6 |
| 19012 | 83 | 8 | 13 | 19 | 1569 |
| 5467 | 27 | 0 | 0 | 10 | 11 |
| 320 | 16 | 0 | 0 | 10 | 11 |

Table 2: Average Statistics

Please note the following:
1. The high number of views remain consistent for the author of popular videos irrespective of the topic of the embedded videos
2. The Likes to Dislike Ratio remains high consistent with point umber 1
3. The Comments/Views ratio indicate that despite a video is generally liked by the viewers, very few will actually go ahead and write a comment concerning the video
4. For average of all the results,
    a. The number of embedded videos was 5 videos
    b. The number of comments go down as the depth of the embedded link increases
    c. The number of comments overall stay quite low regardless of the likes/ dislikes ratio
    d. The likes/dislikes ratio remained favorable over about 47 videos

From the above we infer that viewers do not appear to be affected by the clickbait model if they view the material positively. Given the strength of few videos, an author can attract lots of visitor to his/her channel using the clickbait model in the educational videos environment – an exact opposite behavior of the yellow journalism connotation. The results presented in [17] are consistent with what we have found and we believe that the educational videos do not seem to conform to the traditional negative connotation associated with yellow journalism.

## 5. Conclusion and Future Work

In this paper, we have addressed the existence of clickbait model in the educational videos environment. Furthermore, we believe that we have established enough evidence to Form a basis that the viewers in such an environment display opposite behavior to the negative connotation attached to the clickbait/yellow journalism. The videos we evaluated had both a high viewership and high like count even though the scientific content contained by the videos was not independently verified by experts in the field. We would like to explore the following in the future work:

1. Perform the above experiments on a larger dataset
2. Confirm the high like/dislike ratio by performing sentiment analysis using NLP techniques
3. Repeat the experiment with videos that do not have a high viewership or like/dislike ratio and see if the clickbait model exists in such videos
4. Compare the results of English videos to those in another language to see whether the behavior is consistent across cultures